
\magnification=1200
\hsize=31pc
\vsize=55 truepc
\baselineskip=26 truept
\hfuzz=2pt
\vfuzz=4pt
\pretolerance=5000
\tolerance=5000
 \parskip=0pt plus 1pt
\parindent=16pt

\def\avd#1{\overline{#1}}

\def\th{\rm th}

\def\eqx{\global\advance\equnum by 1
(\the\secnum.\the\equnum)}
\def\eq{\global\advance\equnum by 1
\eqno(\the\secnum.\the\equnum)}

\def\section#1{\global\advance\secnum by 1 \equnum=0 \subsecnum=0
\subsubsecnum=0
\goodbreak\bigskip\bigskip\leftline{\bf\the\secnum. #1}
\nobreak\smallskip\noindent}

\def\subsection#1{\global\subsubsecnum=0\advance\subsecnum by 1
\goodbreak\bigskip\leftline{\rm\the\secnum.\the\subsecnum.
{\it #1}}\nobreak\smallskip\noindent}

\def\subsubsection#1{\advance \subsubsecnum by 1
\medbreak\vskip 12pt{
\noindent{\the\secnum.\the\subsecnum.\the\subsubsecnum.
\it #1.\ }}}

\def\beq{$$}
\def\eeq{\eq$$}

\def\ref{\smallskip\global\advance\refnum by 1 \item{[\the\refnum]}}

\def\QQ{{\rlap {\raise 0.4ex \hbox{$\scriptscriptstyle |$}}
\hskip -0.2em{\rm Q}}}

\def\CC{{ \rlap {\raise 0.4ex \hbox{$\scriptscriptstyle |$}}
\hskip -0.2em{\rm C}}}

\def\frac#1#2{{#1 \over #2}}
\def\i{\ifmmode{\rm i}\else\char"10\fi}

\newcount\equnum
\newcount\secnum
\newcount\subsecnum
\newcount\subsubsecnum
\newcount\fignum
\newcount\refnum

\vskip 1.truecm\noindent
\centerline{\bf Constrained annealing for spin glasses}
\vskip 1.4truecm\noindent
\centerline{Giovanni Paladin$^1$, Michele Pasquini$^1$, and Maurizio Serva$^2$}
\vskip .5truecm
\centerline{\it $^{1}$Dipartimento di Fisica,  Universit\`a dell'Aquila}
\centerline{\it I-67100 Coppito, L'Aquila, Italy}
\vskip .4truecm
\centerline{\it $^{2 }$Dipartimento di Matematica, Universit\`a dell'Aquila}
\centerline{\it I-67100 Coppito, L'Aquila, Italy}
\vskip 1.6truecm
\centerline{ABSTRACT}
\vskip .4truecm
 The quenched free energy of spin glasses can be estimated
 by means  of annealed averages
  where the frustration or other self-averaging variables of disorder
 are constrained to their average value.
We discuss the case of $d$-dimensional Ising models with
random nearest neighbour  coupling, and for
 $\pm J$ spin glasses
   we introduce a new method
 to obtain constrained annealed averages without
 recurring to  Lagrange multipliers.
 It requires to perform quenched averages either on small volumes
 in an analytic way,   or on finite size strips
 via products of random transfer matrices.
We thus give a sequence of converging
 lower bounds for the quenched free energy of $2d$ spin glasses.
\hfill\break
\noindent
PACS NUMBERS: 05.50.+q, 02.50.+s, 75.10.N
\vfill\eject
\section{Introduction}%
Disordered systems
are made of random variables which relax
 to equilibrium on very different time scales.
For instance in the standard models of
 spin glass, the spins arrange themselves to minimize the free energy
   while the random couplings are assumed to be
 frozen in their initial values, since their evolution time is much longer.
As a consequence, one should consider different kinds of averages:
 quenched averages for the frozen variables, and annealed averages
 for the fast variables. However, it is much easier to
 compute annealed averages, which give lower bounds of the
 quenched free energy, the physically relevant quantity.
 These bounds  are often very poor estimates,
 since in annealed averages the main contribution
 comes from a set of disorder realizations
which have zero probability in the thermodynamic limit. In those
 realizations, the quenched random variables are arranged in order to
  minimize the free energy, at difference  with a typical sample
 of the system  where they are frozen in  a given realization.

The idea of constrained annealing is to perform annealed averages
 where the random variables are partly frozen by the requirement
 of satisfying appropriate constraints,
 related to some disorder self-averaging variables.
 The standard way  to impose a constraint
 is by a Lagrange multiplier. By that method, the constrained
 annealed average allows one to introduce
 a sort of Gibbs thermodynamic potential,
 depending on the temperature and on the Lagrange multipliers
 which play the role of the chemical potential
 in ordinary statistical mechanics [1].  In the thermodynamic limit,
 the  values of the multipliers that maximize the Gibbs potential
 are those ones that
 select the realizations with a correct value of
the disorder intensive variables,
 and,  at the same time, that minimize the difference between
 the mean free energy density and its annealed approximations.

Constrained annealing for disordered
 systems has been proposed and applied in particular contexts
  by many authors (see [2], [3], [4], [5], [6]).
 A general formulation of the method and its physical meaning
 can be found in [1].

The purpose of this paper is to show that it is possible to perform
constrained annealed averages  without Lagrange multipliers,
 in $d$-dimensional spin glasses with dichotomic random couplings
 in terms of a quenched average over a finite size sub-system .
  This is often  possible either in an analytic or in a numeric way.
 Using our method, we are able to obtain very good lower bounds
 of the quenched free energy of two dimensional spin glasses.

The summary of the paper is the following:

In section 2, we recall some basic facts about spin glasses.
 We also classify the set of the constraints necessary
 to recover the quenched free energy by annealed averages.
In section 3,  we study the Ising model with couplings
 which can assume the values $J_{ij}=\pm 1$ with equal probability
( the so-called  $\pm J$ spin glass).
In section 4, we introduce our method to obtain
 constrained annealed averages without recurring to Lagrange
 multipliers in the $\pm J$ spin glass.
In section 5, we apply the method to loops of infinite length
 using the Lyapunov exponent of product of random transfer matrices.
We are thus able to  get accurate estimates of the free energy in two
 dimensions.

\section{Constraints for $d$-dimensional spin glasses}%

Spin glasses are the challenge of constrained annealing.
 In that case, the relevant variables to be constrained
 should be related to frustration [7],
  as first remarked by Toulouse and Vannimenus
 in 1976 [5]. As far as we know,
 there are only Montecarlo simulations, see e.g. [8],  [9] ,
 which  used that idea while
  no analytic calculations have been performed on that  ground.
 However, it is possible to get
analytic results  in dimension $d \ge 2$ with constraints
 connected to frustration.
Let us consider $d$-dimensional Ising models
 with nearest neighbour couplings $J_{ij}$ which are
independent identically distributed (i.i.d.) random variables,
 in absence of external magnetic field.
The partition function is
\beq
Z_N =  \sum_{ \{ \sigma \} }  \prod_{i,j}
\exp \{\beta J_{ij} \sigma_i \sigma_j \}
\eeq
where the product runs over $i,j$ which are nearest neighbours.
Using standard relations  of the hyperbolic functions, (2.1) becomes
\beq
Z_N =  \sum_{ \{ \sigma \} }  \prod_{i,j} \cosh(\beta J_{ij})
 \  [ \, 1+  \sigma_i \sigma_j \tanh(\beta J_{ij}) \, ]
\eeq
This form is useful since shows
 that the non-trivial part of the partition function is
 given by
$
 \prod_{i,j}  [1+  \sigma_i \sigma_j \tanh(\beta J_{ij})]
$.
A typical term in that product is $\sigma_i \sigma_j  \sigma_j \sigma_l \cdots
\tanh(\beta  J_{ij}) \, \tanh (\beta J_{jl}) \cdots$,
 where the bonds $\{i,j\} $, $\{j,l\}$ $ \cdots $
 can form either a closed or an open line on the lattice.
If the bonds  form an open line,
 at least two distinct spin variables survive as factor of the product
 of the various  $\tanh(\beta J_{ij} )$, and the related term
 gives a zero contribution to the partition function,
 after performing the sum over the $\sigma$ configurations.
 As a consequence,  one has to consider only
 the terms corresponding to bonds which  form
 a  closed line  (where the product of the $\sigma$'s
 is equal to unity since they appear twice) so that (2.2) can be written
 as a sum over all the possible loops $L_s^{(r)}$,
where $r \ge 4$ indicates the length of the loop ($r$ has to be even)
 and $s$ labels different loops with the same length:
\beq
Z_N=e^{ \sum_{i,j} \ln \cosh (\beta J_{ij} ) } \ 2^N \
 \left( 1+ \ \sum_{ \{ L_s^{(r)} \} }  \ \prod_{ \{i,j\}   \in L_s^{(r)} }
  \tanh (\beta J_{ij}) \right)
\eeq
It follows that in a system  with a finite number $N$ of spins, the
  free energy of a disorder realization  is given by
\beq
 y_N\equiv - {1\over \beta  N} \ln Z_N = - {1 \over \beta} \left( \
   B + \widetilde{F}_N(t^{(4)}, \cdots, t^{(R)}) \right)
\eeq
where $R$ is the maximum possible length of a loop
 in the system ($R < const \, N$), and
$$
B = \ln 2 \, + \, {1\over N} \sum_{i,j} \ln \cosh (\beta J_{ij} )
\eqno(2.4a)
$$
and
$$
\widetilde{F}_N(t^{(4)}, \cdots, t^{(R)})
 =
 {1\over N}
 \ln \left( 1 + \sum_{\{ L_s^{(r)}  \} } \  t_s^{(r)} \right)
\eqno(2.4b)
$$
This function depends on the new set of variables
$$
 t_s^{(r)}
=\prod_{ \{i,j\}  \in L_s^{(r)}} \tanh(\beta J_{ij})
\eqno(2.4c)
$$
given by products which run on the loops $L_s^{(r)}$,   i.e.
plaquettes of larger and larger  perimeter at increasing $r$.
For instance,  the loop $L^{(4)}$ is the elementary plaquette of four bonds.
 Let us stress that a loop ${L}^{(r)}$ can be built up
 as a set of two (or more)  distinct and separate loops
  ${L}^{(r_1)}$ and  ${L}^{(r_2)}$   with $r_1+r_2=r$.
 In the following we often label a coupling  by only one label
 instead of two ones, i.e. $J_{kl}$ is denoted by $J_k$,
since there is no risk of ambiguity on a closed line.

As the number of the nearest neighbour couplings
  in a hypercubic $d$-dimensional lattice is $d N$,
 in the thermodynamic limit  the quenched free energy  reads
\beq
f\equiv \lim_{N \to \infty} \avd{y_N}
 =-{1\over \beta} \ [ \ d \ \avd{\ln\cosh(\beta J_{ij}) }
 + F + \ln 2 \ ]
\eeq
where
\beq
F =
\lim_{N \to \infty} {1\over N} \,  \avd{
 \ln  [ \  1 + \sum_{\{ L_s^{(r)}  \} }
 t_s^{(r)} \ ]
 }
\eeq
 The reduction of the partition function (2.1)
 into (2.3) is the procedure used
  in the Kac and Ward solution of
the $2d$ Ising model  without disorder.

 Eq (2.5) is fundamental since it shows that
 the quenched free energy depends only on the distributions of the variables
 $t^{(r)}$, apart the mean value $\avd{\ln\cosh(\beta J_{ij}) }$.
 The relevant variables to be fixed to their mean value
  in a constrained annealed average of the free energy
 are related, with increasing difficulty
\item{1)} $\ln \cosh \beta J_{ij}$
\item{2)} $t^{(4)}= \tanh(\beta J_{1}) \tanh(\beta J_{2})
 \tanh(\beta J_{3}) \tanh(\beta J_{4})$
\item{3)} $t^{(r)}$ with $r\ge 6$ (considering only connected loops).

 Fixing all the constraints corresponding to those
 quantities in an annealed average is equivalent to
 solve the system with quenched disorder.
This can be understood
 noting that in (2.6) we have a sum over all the possible  $t_s^{(r)}$.
 Because of the law of the large numbers the rescaled sum
 of the $t$-variables corresponding to the same
 topological kind of loops self-averages to
 its quenched mean value.
 Moreover, in the appendix we shall prove that
 this result holds even in a finite volume system
for a $\pm J$ spin glass:
 the quenched free energy can be obtained by a minimization
 of the logarithm of the annealed average of a
 Grand-partition function
 with respect to the Lagrange multipliers related
 to the variables $t_s^{(r)}$ over all the possible loops
 of the finite system.  For a generic distribution
 of the couplings, an extension of this statement
is possible.

In one dimension there are no loops, so that
 the constraint related to point 1)
  is sufficient to obtain the exact solution.
In general, we expect that many physical features
of a spin glass can be reproduced already by considering
 the first two points.

The case of random couplings $J_{ij}=\pm 1$
 with equal probability  is particularly simple in this respect,
 since the variable  $\ln \cosh \beta J_{ij}$ is a constant
 and thus can be ignored.
 Moreover, one immediately sees that (2.4c) becomes
\beq
 t_s^{(r)}=    \tanh^r(\beta ) \, \prod_{\{ i,j \}  \in L_s^{(r)} } J_{ij}
\eeq
implying that the quenched free energy only depends
 on the values of the product of $J_{ij}$
 on closed lines of bonds.
 The non trivial part of the free energy  $y_N$ in (2.4) is thus
\beq
\widetilde{F}_N
 =
 {1\over N}
 \ln (1+x_4 \tanh^4(\beta) +
x_6 \tanh^6(\beta) + \cdots )
\eeq
where the coefficients $\{x_4, x_6,\cdots \}$
 are non-independent  random quantities.
As a consequence, the quenched free energy  is
\beq
f=-{1\over \beta} \left( d \ln \cosh \beta + \ln 2
 + \lim_{N \to \infty}
 {1\over N} \,  \avd{ \ln (1+x_4 \tanh^4(\beta) + x_6 \tanh^6(\beta) + \cdots
)}
\right)
\eeq
In the annealed model, the last non-trivial term of (2.9) vanishes
 since $\avd{x_i}=0$ for all $i$'s (each $x_i$ is a sum of products
 of independent random couplings). In
the limit of high temperature ($\beta \to 0$),
the difference between annealed free energy and quenched free energy
thus is of order $\tanh^4 \beta \sim \beta^4$.
 More important, the simplest
 relevant constraint in the $\pm J$ spin glass is
 the frustration $\widetilde{J}_p$ on the elementary square plaquette
of four couplings $J_1, J_2, J_3, J_4$, that is
\beq
\widetilde{J}_p = - J_{1}J_{2}J_{3} J_{4}
\eeq
 since
 $t^{(4)}= -\widetilde{J}_p \, \tanh^4 \beta $.

\section{$\pm J$ spin glasses}

In this section we consider a $d$-dimensional spin glass
with i.i.d. nearest neighbour couplings
which assume two possible value $J_{ij}=\pm 1$
with equal probability.
{}From (2.9), one sees that the annealed free energy
density is
\beq
f_a = -\frac{1}{\beta} \, \left( \, \ln 2
 + d \, \ln \cosh  \beta \,  \right)
\eeq
In one dimension, there are no loops and $f_a$ is equal to the quenched
free energy density $f$, since $\ln \cosh \beta J_{ij}$
 is a constant,
at difference with
 the gaussian case where that variable is random
 and has to be constrained in order to recover $f$ [10].
In section 2, we have recalled standard results
 which show that
it is  necessary to constrain variables
associated to the loop variable $t^{(4)}$, in order to improve (3.1).
 As first noted by Toulouse and Vannimenus [5], one has to consider
 the intensive variable
\beq
\alpha_N = \frac{1}{N} \sum_{ \{ L_s^{(4)} \}} \widetilde{J}_p
\eeq
with has zero mean value, $\avd{\alpha_N}=0$.
 Let us remark that it is a self-averaging quantity,
 that is $\lim_{N \to \infty} \alpha_N=0$ for almost all disorder
 realizations.
The related constrained annealed average
 can be regarded as the Grand-partition function
$$
\Omega_N(\beta,\mu)\equiv \avd{Z_N  \, e^{- N \, \mu \, \alpha_N} }
$$
In the thermodynamic limit  the Lagrange multiplier $\mu$
that maximizes
$$
g(\beta,\mu)=- \lim_{N \to \infty} {1\over \beta \, N}
 \ln \Omega_N(\beta,\mu)
$$
fixes the frustration on square plaquettes to its mean value.
Moreover it is simple to show the following inequality:
$$
f \ge f_{ac}^{sq} \equiv \, \max_{\mu} \, g(\beta,\mu)
$$

In our case, we have to compute
\beq
\Omega_N (\beta,\mu) =
\avd{\sum_{ \{ \sigma \} }
\exp \{   \sum_{i,j} \beta J_{ij} \sigma_i \sigma_j -
  \sum_{ \{L_s^{(4)} \}} \mu \widetilde{J}_p \}}
\eeq
Taking into account
that the couplings are dichotomic with equal probability, one can
perform the gauge transformation
$J_{ij}  \, \rightarrow \, J_{ij} \sigma_i \sigma_j$
which leaves unchanged the free energy of a disorder realization
 of the system as well as $\widetilde{J}_p$,
 i.e. $\widetilde{J}_p \, \rightarrow \, \widetilde{J}_p$.
As a consequence, (3.3) becomes
\beq
\Omega_N (\beta,\mu)=
 2^N  \ \avd{
\exp \{   \sum_{i,j} \beta J_{ij} -
  \sum_{ \{L_s^{(4)} \} } \mu \widetilde{J}_p \}}
\eeq
This expression (which is meaningful for any
dimension  $d \ge 2$) cannot be computed exactly including all
 the possible square plaquettes $L_s^{(4)}$
of the $d$-dimensional lattice.
 It can be shown that the problem is equivalent
to solve a $d$-dimensional gauge model without disorder,
 whose solution is not known for $d\ge 2$.

However, an exact solution can be achieved if we restrict the sum (3.4)
 only on a part (of order $N$) of the plaquettes.
In other terms the constraint is imposed over a fraction
of the plaquettes, that is we are able to partially fix
the plaquette frustration to its mean value.
As a consequence, the annealed system exhibits a 'residual'
frustration related to the non-constrained plaquettes.

Let us consider the $2d$ model.
It is convenient to limit ourselves to consider
 one half of the plaquettes, which must be chosen in
 such a way that they do not share any coupling,
 as it happens for the black squares of a chessboard.
With this restriction (3.4) contains a sum over
products of $N/2$ independent variables (corresponding
to the $N/2$ black plaquettes) and
the Grand-partition function can be rewritten as:
\beq
\Omega_N (\mu) =
2^N \ \left( \avd{
\exp \{  \beta (J_{1} + J_{2} + J_{3} + J_{4}) +
  \mu J_{1}  J_{2}  J_{3}  J_{4}\}} \right)^{\frac{N}{2}}
\eeq
where $J_{1} \, , \,  J_{2} \, , \,  J_{3} \, , \,  J_{4}$
are the couplings of one of the plaquettes.
After maximizing with respect to $\mu$ the Gibbs potential [1],
 one obtains
\beq
 f_{ac}^{sq}= -\frac{1}{\beta} \, \{ \,
\frac{1}{2}  \ln 2 +
\frac{1}{4} \ln \cosh 2\beta +
\frac{1}{4} \ln (\, 3 \, + \, \cosh 4\beta \, ) \,  \}
\eeq
The functions $f_a$ and $f_{ac}^{sq}$  of the $2d$ spin glass
 are shown in fig 1. At difference with the annealed case
the entropy of the constrained annealed system never
becomes negative.
 We have thus obtained a real qualitative improvement.

It is also interesting to consider
the heat capacity $C$ which can be explicitly computed.
For $\beta \to \infty$ one has
\beq
C \sim \beta^2 e^{-4\beta}
\eeq
The argument of the exponential has been conjectured  in [11] for
the quenched model and it is different from the
exact one-dimension result where the argument  is
$-2\beta$. Note that the annealed system also gives $C \sim \exp(-2\beta)$.

Finally, since we have fixed only one
half of the single plaquette frustration,
it is reasonable to ask what is the residual frustration $\phi_{ac}$.
Since the black plaquettes have zero frustration,
this quantity will be one half of the frustration
of a white plaquette. The four couplings
forming a white plaquette are independent
since they are not coupled by the Lagrange multiplier.
Therefore, the white plaquette frustration can be
written as $-<<J_i>>^4$,
and the global frustration $\phi_{ac}$ reads:
\beq
\phi_{ac}=-\frac{1}{2} <<J_i>>^4
\eeq
where $<<\cdot>>$
is an average on the Gibbs measure associated to
the Grand-partition function (3.3).
One has therefore
\beq
<<J_i>> = -\frac{1}{2}
\frac{\partial ( \beta  g(\beta, \mu^* )  )}{\partial \beta}
\eeq
where $\mu^*$ is the value which maximizes $g(\beta, \mu)$,
and the factor $1/2$ stems from the fact that
the total number
of couplings is $2N$. Since
$$
f_{ac}^{sq} (\beta)= g(\beta, \mu^* (\beta))
$$
and since the derivative with respect to $\mu$ vanishes,
one has
\beq
<<J_i>> = -\frac{1}{2} \frac{d (\beta f_{ac}^{sq} ) }{d\beta}
\eeq
Using the expression (3.6) for $f_{ac}^{sq}$ we
can compute the residual frustration
\beq
\phi_{ac}= -\frac{1}{2} <<J_i>>^4 =
-\frac{1}{2}  \, \left[ \frac{\th (2\beta)}{4} +
\frac{ \th (2\beta)}{4-2 \th^2 (2\beta)} \right]^4
\eeq
At zero temperature
 $\phi_{ac} =3^4 /2^9 \approx 0.158$.
This result indicates the presence of a small
but non vanishing residual frustration.
An analogous expression can be found
from the annealed free energy $f_a$.
In this case all plaquettes can have negative frustration and
one has
\beq
\phi_a = -<<J_i>>^4 =- \th^4(\beta)
\eeq
where the average is in the Gibbs measure generated by the
annealed average. The frustration $ \phi_a(T=0)=-1$ since at zero temperature
 the spins are aligned and the annealed system is
completely ferromagnetic. Both $\phi_a$ and $\phi_{ac}$
are reported in Fig 2.
The fact that in the constrained annealed model
the residual frustration does not vanish indicates that
the free energy estimate could be improved by
further constraints as we shall discuss in the next section.

\section{Constraints without Lagrange multipliers}

As far as we know, up to now constrained annealing has been
 performed only using Lagrange multipliers.
This is a strong limitation, since
in most cases it is too difficult to derive
 analitically an annealed average using more than one constraint.
 The spin glass with dichotomic couplings has
some algebraic features which permit us to introduce an alternative method,
in order to obtain constrained annealed averages
avoiding in this way the problem of the minimization
 of the Lagrange multipliers.

Let us consider a $d$-dimensional lattice of $N$ points:
the total number of the couplings is $d N$, but the system
can be fully described by a lower number of independent random
variables. In fact the non-trivial part of the quenched free energy
 (2.6) is a function of the variables  $\{t_s ^{(r)} \}$,
 see (2.4c) , defined on the closed loops of the lattice.
Indeed,
from a topological point of view
any closed loop $L_s ^{(r)}$ can be thought as the union
of some elementary square plaquettes, say a number $k$,
so that the related variable $t_s ^{(r)}$ can be written as:
\beq
t_s ^{(r)}= \tanh^r (\beta) \ \prod_{i=1}^k \widetilde{J}_p^{(i)}
\eeq
where we have introduced the frustrations
$\widetilde{J_p}=\prod_{i=1}^4 J_i =\pm 1 $ with equal probability
of the $k$ plaquettes (the index $i$ runs over the four sites
of each square). Let us stress that a loop can be either connected or
 disconnected,
i.e. it is not  necessary that the plaquettes are neighbours.
Notice that  the plaquette frustration was previously defined
with a minus sign (see (2.10)),
but in this context we neglect it to simplify the notation.
Moreover, we dot not assume periodic
boundary conditions in this section.

In other terms, we can consider the plaquette frustrations
as the random variables of the system, instead of the couplings,
with no loss of generality. We shall see that the number of
frustrations necessary for a full description of the system is always lower
 than $d N$, the total number of couplings.
 For instance, in 2$d$, there are 2$N$ couplings,
while the number of plaquette frustrations is of order $N$ when
 $N \to \infty$.

However, the $2d$ case exhibits a special feature, since all
the possible plaquette frustrations turn out to be a set of
independent random variables (let us recall that we do not
assume periodic boundary condition).
In the general case $d \ge 3$,  this statement is no longer valid.
Indeed, let us consider a cube in a $d$-dimensional lattice:
a moment of reflection
shows that the product of the related six plaquette frustrations
has to be 1, so that only five frustrations are independent
random variables.
As a consequence, for $d \ge 3$ the change of
random variables from the couplings to the plaquette frustrations
involves only a subset of them, let us say $N^*$,
chosen in such a way that they are
 independent random variables and that they can build up
every possible closed loop on the lattice.

Let us illustrate this change of random variables for a 3$d$ lattice
of $N$ points, where we can introduce three integer coordinates
$\{x, y, z \}$. A square plaquette has three different orientations,
belonging to a $x= const$ plane, or $y=const$, or $z=const$.
In the thermodynamic limit one has $N^{1\over 3}$ planes for each
orientation, and every plane is a set of $N^{2\over 3}$ plaquettes,
for a total number of $3 N$ plaquettes in the whole lattice.

Consider a cube of six plaquettes, with one of them
on the $z=0$ plane. As previously discussed, one of the $\widetilde J$'s
cannot belong to the subset of plaquette frustrations that describe
the whole system, let us say the frustration of the plaquette
on the $z=1$ plane.
Let us now consider the other cube  sharing this `neglected' plaquette.
 By introducing as random variables of the system
the frustrations which are  related to the four plaquettes that
share a coupling with the plaquette on the  $z=1$ plane,
we have to neglect the sixth frustration of the plaquette
on the  $z=2$  plane,
since the product of all the plaquette frustrations of these
two nearest neighbour cubes is fixed to 1 (the presence of the plaquette
on $z=1$ in the product is irrelevant, since it appears twice).
These arguments can be easily repeated, so that the subset of plaquette
frustrations that fully describes the whole system can be found as follows:
one has to consider two distinct orientations and to take into account
all the plaquettes with these orientations, for a total number of $2 N$.
Moreover, the third orientation gives only a plane of $N^{2\over 3}$
plaquettes (the $z=0$ plane in our example), so that
in the thermodynamic limit the number
of independent random frustrations $N^*$
that fully describe the systems grows as $2 N$.

The generalization of this result is immediate: in the thermodynamic limit
the $d$-dimensional lattice of $N$ points (with $d \ge 3$) is built up by
$N \pmatrix{ d \cr 2}$ distinct plaquettes,
since $\pmatrix{ d \cr 2}$ is the number of different orientations
for a plane in the lattice, but only a number $N^* =2 N$
of plaquette frustrations is necessary to describe the system.
In fact the change of random variables involves only the frustrations
of the plaquettes with two well-defined orientations ($2 N$ plaquettes),
together with a plane of plaquettes for each one of the
other orientations, but in the thermodynamic limit
the main contribution to $N^*$ comes from the former term.
 Using this subset of plaquettes,
whose related frustrations
are all independent random variables,
it is possible to build up every possible closed loop on the
$d$-dimensional lattice.

In conclusion the described change of variables  permits us to reduce
the number of independent random variables necessary
to describe the system, from $d N$ couplings to $N^*$ frustrations,
with  $N^*=N$ for {\it d}=2 or $N^*=2 N$ for $d \ge 3$.
In the following, if not differently specified, the term `frustration'
always indicates one
of the $N^*$ random variables of the system.

Recalling (4.1), equation (2.6) becomes:
\beq
F= \lim_{N \to \infty} {1\over N} \
 \avd{
\ln  [ 1 +\sum_{k=1}^{N^*} \ \sum_{i_1 \dots i_k}
\widetilde{J}_p^{(i_1)} \dots \widetilde{J}_p^{(i_k)}
 \tanh^{r(i_1, \dots , i_k)} (\beta) ]
}
\eeq
where one considers all the products of $k \in [1,N^*]$
elementary plaquettes
( the indices  $ i_1 \neq i_2 \neq \dots \neq i_k$ run from 1 to $N^*$)
 and $r(i_1, \dots, i_k)$ is the length of the loop
 built up by the $k$ plaquettes, while the overline represents now
the average over the frustrations.

In the following paragraph, we consider the plaquettes
of the lattice only from a topological point of view, i.e.,
the fact that a frustration of a plaquette is a random variable
of the system (or not) is irrelevant.
At this point, we perform a decomposition
 of the set of plaquettes
   into subsets such that each coupling $J_{ij}$
 belongs to one and only one of these subsets. It follows that two
 distinct subsets can have in common only isolated lattice points.
 For instance, three different decompositions
 of the bonds of a 2$d$ lattice (the black regions)
 are illustrated in fig 3.
 It is worth stressing that, after the decomposition,
 we get a collection of sub-systems
 which do not cover the whole original lattice, e.g.
  in fig 3a  only one half of the 2d lattice is
 covered.
This kind of decomposition
divides the plaquettes into two classes: the `black' ones, which are organized
in groups corresponding to the sub-systems (e.g. the crosses in fig 3b), and
the `white' ones, which do not belong to any sub-system.

Coming back to the frustrations, note that
in general there are many `white' or
`black' plaquettes whose frustration is not a random variable
of the system.

At this point we can derive an upper estimate $F_{ac}$ of the function (4.2),
by treating all the random variables of frustration associated to the
`white' plaquettes as annealed variables, i.e. by  averaging
over these variables only the argument of the logarithm in (4.2),
instead of the logarithm itself:
\beq
F_{ac} = \lim_{N \to \infty} {1\over N}
 \avd{
\ln  [ 1 +\sum_{k=1}^{N^*} \ \sum_{i_1 \dots i_k}
\avd {\widetilde{J}_p^{(i_1)} \dots \widetilde{J}_p^{(i_k)}}^{(w)}
 \tanh^{r(i_1, \dots , i_k)} (\beta) ]
}^{(b)}
\eeq
where $\avd{ \ \bullet \ }^{(w)}$ and $\avd{\ \bullet \ }^{(b)}$
represent the
averages over the frustrations related to, respectively, the `white' and the
`black' squares.
In (4.3) only the terms $\widetilde{J}_p^{(i_1)} \dots \widetilde{J}_p^{(i_k)}$
with all `black' frustrations do not vanish after
performing the `white' average:
\beq
F_{ac} = \lim_{N \to \infty} {1\over N}
 \avd{
\ln  [ 1 +\sum_{k=1}^{N_b} \ \sum_{i_1 \dots i_k}
\widetilde{J}_p^{(i_1)} \dots \widetilde{J}_p^{(i_k)}
 \tanh^{r(i_1 , \dots  , i_k)} (\beta) ]
}^{(b)}
\eeq
where now the indices $i_1, \dots , i_k$ run only over the $N_b$
`black' frustrations.
It should be remarked that after performing the `white'
 average, the surviving loops
 do not connect different sub-systems. In other terms,
 any loop  appearing in the average (4.4) can be decomposed into a
 set of `sub-loops', each of them limited to a single sub-system.
This implies that $\tanh^{r(i_1, \dots , i_k)} (\beta)$ can be factorized,
i.e., $r(i_1, \dots , i_k)$ is the sum of the lengths of the various
`sub-loops'. As a consequence, the whole argument of the logarithm in (4.4)
is factorized among all the sub-systems, and it immediately follows:
\beq
F_{ac} = {d\over n_j} \  \avd{
\ln  [ 1 +\sum_{k=1}^{n_p} \ \sum_{i_1 \dots i_k}
\widetilde{J}_p^{(i_1)} \dots \widetilde{J}_p^{(i_k)}
 \tanh^{r(i_1, \dots , i_k)} (\beta) ]
}
\eeq
where the average is performed over the $n_p$  `black'
plaquette frustrations that belong to a single sub-system, and
where $n_j$ is the number of couplings in the sub-system, so that
$d\over n_j$ represents the total number of sub-systems, rescaled with $N$
in the thermodynamic limit.
Recalling (2.5), from (4.5) the lower estimate $f_{ac}$ of
the quenched free energy yields:
\beq
- \beta f_{ac}= d \ \ln\cosh(\beta) +
{d\over n_j} \ \avd{
\ln  [ 1 +\sum_{k=1}^{n_p} \ \sum_{i_1 \dots i_k}
\widetilde{J}_p^{(i_1)} \dots \widetilde{J}_p^{(i_k)}
 \tanh^{r(i_1, \dots , i_k)} (\beta) ]
} \
 + \ln 2
\eeq
It is preferable to choose a
topological decomposition of the system such that the $n_p$
`black' frustrations fully describe the single sub-system.
In this context $F_{ac}$ is  practically equivalent
to the quenched free energy of the sub-system, a part a factor proportional
to $\ln\cosh \beta$.
If $n_p$ is not too large, the computation of $F_{ac}$ can be performed
analitically or numerically.

The correct multiplicative factors to transform the
 quenched free energy of a sub-system  to the global constrained
 free energy (4.6) can also be obtained
 by a simple argument. Indeed
the partition function $Z_{n_s}$  of a sub-system
 made of $n_s$ spins and $n_j$ couplings
 is the sum of $2^{n_s}$ terms
 which are given by the product of $n_j$ exponentials
 while the global constrained Grand-partition function
 $\Omega_N$  of a system with $N$ spins and $d \, N$
 couplings is the sum of $2^{N}$ terms
 which are given by the product of $d \, N$ exponentials.
In order to compare two quantities of order one, one has
 to write the following equality:
$$
( \, 2^{-N} \Omega_N \, )^{1\over d\,N}=
( \, 2^{-n_s} Z_{n_s} \, )^{1\over n_j}
$$
so that
\beq
f_{ac} = - {1\over \beta} (1-d \ {n_s\over n_j}) \ \ln 2 \ + \
d \ {n_s\over n_j} f_{sub} \, .
\eeq
This formula is completely equivalent to (4.6),
 but has the advantage that
the quenched free energy $f_{sub}$
of the sub-system explicitly appears.
The ground state energy can be estimated by the saddle point method
 when  $\beta \to \infty$ in the quenched average of $Z_{n_s}$:
\beq
E_{ac}(T=0)={d \over n_j} \, \avd{ \max_{\{\sigma\}}  H  (\{J\}) }
\eeq
where the maximum is taken over all the
 $\sigma$ configurations for each disorder realization.
 One can also obtains the residual entropy as
\beq
S_{ac}(T=0)= (1-{n_s\over n_j} d) \, \ln 2 + \avd{\ln deg(\{J\}) }
\eeq
where $deg(\{J\})$ is the number of disorder configurations which
have energy $ H  (\{J\})$ equal to $E_{ac}(T=0)$.

Let us briefly resume the discussion, in order to clarify the meaning
of the result. Using a topological
decomposition of the system into independent
sub-systems, we have derived a lower estimate of the quenched free energy
that depends from the quenched free energy of the single
sub-system. As discussed in the appendix,
 the quenched   free energy of a sub-system can be obtained by a
minimization of an annealed average of the Grand-partition function
  with Lagrange multipliers
over all its possible loops. Using our procedure,
we are able to get the annealed average  $f_{ac}$
 of the global system where
the constraints are imposed over all the possible rescaled sums of
frustration on
all the loops  that appear in the various sub-systems.

To illustrate our method, we apply it in the case of sect 3,
 i.e. the  decomposition of  a $2d$ spin glass in `white' and
`black' square plaquettes, like in a chessboard, see fig  3a.
In this case all the $N$ plaquette frustrations represent
a set of independent random variables, i.e. $N^*=N$. One half of them
is related to `white' plaquettes, so that they are
treated as annealed variables, while the other half related to
`black' plaquettes as quenched variables.
We also have $n_j=4$ and $n_p=1$, so that (4.6) becomes:
\beq
- \beta f_{ac}^{sq}= 2 \ \ln\cosh(\beta) +
 {1\over 4} \ln  [ 1 - \tanh^{8} (\beta) ] + \ln 2
\eeq
It is easy to check that after trivial algebraic manipulations one
gets again (3.6).

The ground state energy and the zero temperature entropy can be
directly computed from (4.10), and respectively are:
\beq
E_{ac}^{sq}(T=0)=-1.5
\qquad
S_{ac}^{sq}(T=0)=0
\eeq

The method of constrained annealed averages without multipliers
  can be applied to non-elementary sub-systems
in an easy numerical way, since one has not to introduce a set of
 corresponding Lagrange multipliers.
In order to improve (4.10) and (4.11), we
 have considered a partition of the $2d$ lattice into independent
 crosses of $n_p=5$ square plaquettes, as shown in fig 3b.
We thus obtain a constrained annealed average
 where all the relevant variables on the non connected loops
 inside the crosses are frozen.
Since $n_j=16$, from (4.6) the free energy is easily evaluated,
and  is shown in fig 1.
 The ground state energy and the residual entropy are respectively
\beq
E_{ac}^{cr}(T=0)=-1.484375 \qquad S_{ac}^{cr}(T=0)=0.00882...
\eeq

The next step is to considered the
elongated cross of the type shown in fig 3c ($n_p=8$ and $n_j=24$).
In this case we get
\beq
E_{ac}^{ecr}(T=0)=-1.477865 \qquad S_{ac}^{ecr}(T=0)=0.0130...
\eeq
At increasing the size of the subsets of the decomposition,
 the convergence to the quenched ground state energy
$E_0$ is rather slow  as the numerical result  of [11] gives
$E_0 =-1.404 \pm 0.002$  and   $S_0=0.075 \pm 0.004$).
 However, the main qualitative  feature
(positive residual entropy) is reproduced
 by our approximations.

We have
also applied our technique to the $\pm J$ spin glass
in three dimensions. From a topological point of view, a $3d$ lattice
with $N$ points can be thought as the union of distinct cubes
distributed in such a way that
two of them  have only one lattice point in common
(the 3$d$ equivalent of the 2$d$ chessboard).
In the thermodynamic limit their number is ${1\over 4} N$.
Let us stress that in this case the six plaquettes of each cube
represent the `black' regions for a total of ${3\over 2} N$,
while there are ${3\over 2} N$ `white' plaquettes between cubes.

It is easy to realize that there exists a different, but equivalent,
change of random variables with respect to the one previously described,
which involves  $n_p=5$ `black' plaquette frustrations
for each cube (the neglected plaquette must be always the same),
together with
other plaquette frustrations in the `white' areas, for a total of $2 N$
independent random variables.
After performing the `white' annealed average, the problem
 is reduced
to the calculation of the quenched free energy of a single cube.
In this case we obtain a constrained annealed average of the free energy of
the global system $f_{ac}^{cube}$,
where the constraints are imposed over the products
of the $J_{ij}$'s on all the closed loops of bonds in the ${1\over 4} N$
independent cubes.
The resulting $f_{ac}^{cube}$ has a very long
 analytic expression and is plotted in fig 4,
where it is also  drawn for comparison the annealed free energy.
 Unfortunately, the residual entropy is still negative, thus indicating that
 in three dimensions more constraints are necessary to get
 a fair approximation of the quenched system at low temperature.
 However, the free energy should be a non-decreasing function of the
temperature,
 so that the ground state energy can be  estimated
 as the supremum of an annealed approximation although the free energy has
  negative derivative at low temperature  [5].
 Our constraints on all the loops inside the alternated cubes
 allows to estimate
 $E_0 (d=3) \ge -1.917$ which improves the lower bounds
given  by the supremum
 of the annealed free energy, $E_0 \ge -1.956$

\section{Constrained annealed averages on infinite loops}
The method introduced in the previous section  could allow us to
 obtain very accurate estimates (lower bounds)
of the quenched free energy.
Indeed the main limitation we have met is that
 the number of plaquettes $n_p$
 of the sub-system should be not too large in order
 to perform an analytic calculation of its quenched free energy.
 For instance in $2d$ we have stopped
 the estimates  at the elongated cross with $n_p=8$ plaquettes.
 However, we can consider infinite loops by estimating
 the free energy of  sub-systems of infinite size only in one direction,
  via the Lyapunov exponent of the product  of random transfer matrices
 [12].

To simplify, let us consider again the two dimensional Ising model
with  $J_{ij}=\pm 1$ with equal probability,
 although our discussion can be extended to three and higher dimensions.
The idea is  to find an independent decomposition of the  $2d$ lattice
 in strips of size $L$. We then compute the quenched free energy of
 the strip as the Lyapunov exponent
 of the infinite product of random transfer matrices of size
 $2^L \times 2^L$. In such a way,  we automatically obtain
 the annealed average where the products of $J_{ij}$'s
 on all the possible loops inside the strip are constrained to their
 quenched mean value.  In particular, some of the loops are of infinite
size in one direction.
 Our proposal requires
  a numerical calculation of the Lyapunov exponent, but it
 is superior to the direct application of the transfer matrix method [13],
 since it allows one to obtain lower bounds of the free energy
 which become more and more accurate
 at increasing the size $L$ of the strips, as a consequence of the
 standard inequalities satisfied by the constrained annealed averages.

 In order to reduce the number of random variables, it is convenient
to perform a gauge transformation,
before computing the free energy of the spin glass.
 In two dimensions, we have chosen
a gauge transformation which map the original system into a new one where
 the horizontal couplings are i.i.d.
 random variables   ($J_{ij}=\pm 1$ with equal probability)
 and the vertical couplings are positive and constant ($J_{ij}=1$).
 We  consider
 strips of $L$ spins which are
parallel to the bisectrix of the lattice
 as shown in fig 5, in order to obtain a convenient
 independent decomposition
 of the lattice with the properties discussed at the beginning of  sect 4.
 The basic cell of the strip is formed by three layers,
 the first and the third ones of length $L$ and the intermediate layer of
 length $L+1$.
 We denote by $\sigma$ the spins of the first layer,
$\xi$ the spins of the second layer and $\eta$ the spins of the third layer.
The particular  form of the strip allows one
 to perform a preliminary analytic integration over the spins of
 intermediate layers.
 For instance, the case $L=3$
 is illustrated in fig 5.
 The partition function of that strip
 can be obtained by a product of transfer matrices
 between the first and the third spin layers:
\beq
\widetilde{Z}_N(L=3)=
Tr \ \prod_{i=1}^N {\bf T}(i;J_1,\cdots, J_6 \, )
\eeq
where  there are $2^6$  different  random transfer matrix  $\bf{T}$.
As the Boltzmann factor $\exp(-\beta H)$
 of three consecutive layers of the
 $L=3$ strip is
$$ \exp(\sigma_1 \xi_1+\sigma_2 \xi_2+\sigma_3 \xi_3) \,
 \exp(J_1 \sigma_1 \xi_2+J_2 \sigma_2 \xi_3+J_3 \sigma_3 \xi_4) \,
$$
$$
 \exp(\eta_1 \xi_2+\eta_2 \xi_3+\eta_3 \xi_4) \,
 \exp(J_4 \eta_1 \xi_1+J_5 \eta_2 \xi_2+J_6 \eta_3 \xi_3)
$$
one immediately
 sees that, after integrating out the intermediate lay (the four $\xi$ spins),
the elements of the $8 \times 8$  transfer matrices  are
$$
T_{\sigma_1,\sigma_2,\sigma_3, \eta_1, \eta_2, \eta_3}=
$$
\beq
\cosh(\sigma_1+J_4 \eta_1)
\cosh(J_1 \sigma_1+J_5 \eta_2 +\sigma_2+\eta_1)
\cosh(J_2 \sigma_2+J_6 \eta_3 +\sigma_3+\eta_2)
\cosh(J_3 \sigma_3+\eta_3)
\eeq
The extension to a generic value of $L$ is straightforward,
 and one has to deal with the product of $2^{2L}$ independent random matrices
 of size $2^L \times 2^L$.

The Lyapunov exponent $\lambda$ of the product of the transfer matrices
is related to the quenched free energy $f_{strip}^{(L)}$ [12] by:
\beq
\lambda=- n_s \beta f_{strip}^{(L)}
\eeq
A moment of reflection shows that there are $n_s=2L+1$ spins and
$n_j=4 L$ couplings, so that from (4.7) one obtains
the constrained annealed free energy $f_{ac}^{(L)}$ of the total system:
\beq
-\beta f_{ac}^{(L)}={\lambda - \ln 2  \over2 \, L}
\eeq
The constrained
 free energy as function of the temperature for $L=1$, $L=2$, $L=3$
 and $L=8$ is shown in fig 6.
 It is evident that the difference between
 the constrained annealed free energy $f_{ac}^{(L)}$ and the quenched one
 increases at lowering the temperature, since
 the frustration effect becomes more and more important.
 In two dimensions at varying the size of the transversal length $L$
we found a monotonous convergent sequence of approximations
from below to the ground state energy and to
 the residual entropy:
$$L=1  \qquad E_0=-1.500 \qquad S_0=0.00$$
$$L=2  \qquad E_0=-1.464 \qquad S_0=0.02$$
$$L=3 \qquad E_0=-1.448 \qquad S_0=0.03$$
$$L=4  \qquad E_0=-1.438  \qquad S_0=0.04$$
$$L=5  \qquad E_0=-1.429 \qquad S_0=0.05$$
$$L=6  \qquad E_0=-1.425 \qquad S_0=0.05$$
$$L=7  \qquad E_0=-1.423 \qquad S_0=0.05$$
$$L=8  \qquad E_0=-1.421 \qquad S_0=0.05$$
$$L=9  \qquad E_0=-1.419\qquad S_0=0.06$$

In conclusion, we have shown that the constrained annealing
in $\pm J$ spin glasses can be performed
 without Lagrange multipliers, by performing quenched averages
 on appropriate sub-systems.
 However, it is an open problem to extend it to other coupling distributions
 such as the gaussian distribution.

\bigskip
\bigskip
{\bf Acknowledgements}
 \medskip
GP and MS acknowledge the financial support
 ({\it Iniziativa Specifica} FI3) of the I.N.F.N.,
  National Laboratories  of Gran Sasso,
{\it Gruppo collegato dell'Aquila}.
\vfill\eject

\magnification 1200
\hsize=31pc
\vsize=55 truepc
\baselineskip=26 truept
\hfuzz=2pt
\vfuzz=4pt
\pretolerance=5000
	\tolerance=5000
 \parskip=0pt plus 1pt
\parindent=16pt

\def\avd#1{\overline{#1}}
\def\a1{\alpha_1 (x)}
\def\ai{\alpha_i (x)}

\def\an{\alpha_n (x)}

\vfill\eject
{\noindent
\bf Appendix: How many Lagrange multipliers for the quenched free energy?}

Strictly speaking, the method of Lagrange multipliers, as formulated
in [1] shows that computing
an annealed average of the partition function with some constraints
 which hold  in the thermodynamic limit
is equivalent to determining  the maximum of an appropriate
Gibbs potential $g(\beta,\mu)$ with
 respect to $\mu$.
Let us note that in general $\mu$ indicates
   a whole set of Lagrange multipliers.
Actually, this method suggests a clever way to calculate a quenched average,
 by finding the minima of annealed models with Lagrange multipliers  even
 in finite volume systems.  However, in this case,
the Lagrange multipliers do not constrain the variables $\alpha_N$
 to their mean value.
Only in the thermodynamic limit these constraints are achieved by a saddle
point estimate of the annealed average of the Grand-partition function.
In both cases (finite or infinite volume systems) the most interesting
question is to establish how many and what quantities are necessary
to reproduce a quenched average.
In this appendix we give an explicit  answer.
We limit our considerations to the case of a single random variable $x$,
since the following arguments can be easily extended to
many or infinite variables.
Let us suppose that $x$ is defined over a set $X$ with a probability density
$p(x)$, and let $Z(x) \ge 0$ be a function of $x$ such that the quenched
average $\avd{\ln Z(x)}$ exists finite.
One has the following theorem:

\bf Theorem: \rm Assume there is a set of $n$ functions
$\{ \a1, \dots , \an \}$ and a set of $n+1$ constants $\{ c_0, \dots , c_n \}$
such that:
$$
\ln Z(x) \ = \ c_0 \ + \ \sum_{i=1}^n c_i \, \ai
\eqno(A1)
$$
on $X$. Then the quenched average $\avd{\ln Z(x)}$ is equal to:
$$
\avd{\ln Z(x)} \ =\  \min_{ \mu_1, \dots , \mu_n }
\ \ln \ \avd{ \, Z(x) \ \exp \left[ - \sum_{i=1}^n \mu_i
\left(\ai - \avd{\ai} \right) \right]}
\eqno(A2)
$$
and the minimum is reached for $\{ \mu_i=c_i\}_{i=1, \dots , n}$.

\it Proof: \rm Recalling (A1), from a direct inspection it follows that
the right hand side of equation (A2) is equal to $\avd{\ln Z(x)}$ for
$\{ \mu_i=c_i\}_{i=1, \dots , n}$. By virtue of the convexity
of the logarithm the proof is completed.
Let us stress that, in general, there is an infinite number $n$
of functions $\{ \ai \}$.

This result, i.e. that a quenched average is
formally equivalent to a variational annealed problem, was first
established by Morita [2], using, indeed, a different
variational approach with respect to relation (A2).

This simple theorem can be usefully applied in the case
of finite systems with dichotomic disorder.
Let us consider, for instance, a 2$d$ Ising spin glass with
nearest neighbours random couplings $\{ J_{ij}=\pm 1 \}$. As previously seen,
the partition function $Z_N$ of the system of $N$ spins depends only on
the plaquette frustrations $\{ \widetilde J_p ^{(i)} \}$,
which are a set of independent dichotomic
($\pm 1$) random variables (no boundary conditions are assumed).
{}From elementary properties of the dichotomic variables, the free energy
$y_N =- {1 \over \beta N} \ln Z_N $
 of a realization of the system can be written as:
$$
y_N =
\ c_0 \ + \ c_1 \ {1\over N} \sum_{L^{(4)}}  J_p^{(i)}
\ + \ c_2 \ {1\over N} \sum_{L^{(6)}}  J_p^{(i)} J_p^{(j)} \
+ \ c_3 \ {1\over N} \sum_{L^{(8)}}  J_p^{(i)} J_p^{(j)} \ +
$$
$$
+ \ c_4 \ {1\over N} \sum_{L^{(8)}}  J_p^{(i)} J_p^{(j)} J_p^{(k)}\ +
+ \ c_5 \ {1\over N} \sum_{L^{(10)}}  J_p^{(i)} J_p^{(j)} J_p^{(k)}\ +
+ \ c_6 \ {1\over N} \sum_{L^{(12)}}  J_p^{(i)} J_p^{(j)} J_p^{(k)}\ + \  \dots
$$
where there are different terms with the same number of plaquettes in the
product,
according to the length of the resulting loop built up by these plaquettes.
For instance, there are  three
terms by the sums of products of three plaquettes, corresponding to
loops of length 8, 10 and 12.
Thus the quenched free energy can be recovered with an annealed average
of type (A2) where every possible function
$$
{1\over N} \sum_{L^{r(i_1, \dots , i_k)}}  J_p^{(i_1)} \dots  J_p^{(i_k)}
\eqno(A3)
$$
appears with its Lagrange multiplier.
In the thermodynamic limit ($N \to \infty$) this means that the quenched
free energy is equal to an annealed free energy where
the frustration computed along every kind of closed loop on the lattice
is constrained to its mean value.

This last statement keeps its validity
 for a generic distribution of the couplings,
taking into account that the frustrations along the closed loops
 should be
substituted for the more generic variables
$\{ t^{(r)}_s \}$ defined by (2.7). In other terms, the general quantity
to constrain (corresponding to (A3) for the dichotomic case) is:
$$
{1\over N} \sum_{L^{r)}}  t^{(r)}_s
\eqno(A4)
$$
The proof can be achieved by arguments derived from the large number law.
On the contrary, for a finite size system it is not sufficient
to perform an annealed average with constraints over all
the quantities (A4) to recover the quenched free energy, at difference
with the dichotomic case.
Indeed, the $\pm J$ spin glass has a special feature, i.e., both the
partition function and its logarithm can be expressed in terms
of only the quantities (A3), by virtue of the dichotomic properties
of the random variables, while for a generic distribution
 only the partition function can be written as
sum of the quantities (A4).
However, in principle one can get an expression of the logarithm of
the partition function recurring to power series of the (A4)'s,
and, as a consequence, one can find
all the quantities to constrain in order to obtain
the quenched free energy for a finite volume system.

\vfill\eject
\noindent
{\bf References}
\bigskip
\bigskip
\bigskip
\item{[1]}
M. Serva and G. Paladin, {\it Gibbs
thermodynamical potentials for disordered systems},
Phys. Rev. Lett. {\bf 70},  105 (1993)

\bigskip
\item{ [2]}
T. Morita,
{\it Statistical mechanics
of quenched solid solutions with applications to
  diluted  alloys},
J. Math. Phys. {\bf 5}, 1401 (1964)

\bigskip
\item{[3]}
R. K\"uhn, D. Grensing, H. Huber,
{\it Grand ensemble solution of a Grand ensemble
 spin glass model},
 Z. Phys. B {\bf 63}, 447 (1986)

\bigskip
\item{[4]}
M. F. Thorpe and D. Beeman,
{\it Thermodynamics of an Ising model with random exchange interactions},
Phys. Rev. B {\bf 14}, 188 (1976)

\bigskip
\item{ [5]}
G. Toulouse and J. Vannimenus,
{\it On the connection between spin glasses and
gauge field theories},
Phys. Rep. {\bf 67}, 47 (1980)

\bigskip
\item{[6]}
J. Deutsch and G. Paladin,
{\it The product of random matrices in a microcanonical ensemble}
 Phys. Rev. Lett. {\bf 62}, 695 (1988)

\bigskip
\item{[7]}
G. Toulouse,
{\it  Theory of the frustration effect in spin glasses},
Commun. Phys. {\bf 2}, 115 (1977)

\bigskip
\item{ [8]}
G. Aeppli and G. Bhanot,
{\it Ising spin gauge theory and upper marginal dimensionality for spin
glasses},
J.  Phys. {\bf C14}, L593 (1981)

\bigskip
\item{ [9]}
G. Bhanot and M. Creutz,
{\it Ising gauge theory at negative temperatures  and spin glasses},
Phys. Rev.   {\bf B22}, 3370 (1980)

\bigskip
\item{[10]}
S. Scarlatti, M. Serva and M. Pasquini, {\it
Large deviations for Ising spin glasses with constrained disorder},
J. Stat. Phys. (submitted)

\bigskip
\item{[11]}
L. Saul and M. Kardar,
{\it Exact integer algorithm for the two-dimensional
Ising spin glass},
Phys. Rev. E {\bf 48} (1993) 48

\bigskip
\item{[12]}
A. Crisanti, G. Paladin and A. Vulpiani
{\it Products of random matrices in statistical physics}
(Series in Solid State Sciences 104,  Springer-Verlag) (1993)

\bigskip
\item{[13]}
H.-F. Cheung and W. McMillan
{\it Equilibrium properties of the two dimensional random $\pm J$
 Ising model},
 J. Phys. {\bf  C16}, 7027 (1983)

\vfill\eject
\vskip 0.8truecm
\centerline {\bf Figure Captions}
\vskip 0.5truecm
\noindent

\item{Fig. 1}
$\pm J$ spin glass in $2d$:
the annealed free energy $f_a$ (dashed line) and the
constrained free energies $f_{ac}^{sq}$ (full line)
and $f_{ac}^{cr}$ (dot-dashed line) versus temperature $T=1/\beta$.
The constraints are imposed, respectively, on $N/2$ independent plaquettes
and on $N/8$ independent crosses of five plaquettes.

\medskip
\item{Fig. 2}
$\pm J$ spin glass in $2d$:
residual frustration  $\phi_a$ (3.12)
for the  annealed model
 and $\phi_{ac}$ (3.11) for
the constrained model. The constraint is imposed on $N/2$
independent plaquettes.

\medskip
\item{Fig. 3}
Ising model with nearest neighbour interactions
on a $2d$ lattice: decompositions of the set of the couplings in independent
subsets (the black areas): a) squares, b) crosses and  c) elongated crosses.
\medskip
\item{Fig. 4}
$\pm J$ spin glass in $3d$:
  the annealed free energy $f_a$ (dashed line) and
constrained free energy $f_{ac}^{cube}$ (full line)  versus $T=1/\beta$.
\medskip
\item{Fig. 5}
Ising model with nearest neighbour interactions on a $2d$ lattice:
\item \item a) decompositions of the set of the couplings into independent
subsets of infinite area (the black strips) in the case $L=3$.
\item \item b) the basic cell of the strip ($L=3$) is formed by three layers:
$\sigma$, $\xi$ and $\eta$ spins; after the gauge transformation the vertical
couplings are fixed ($J_{ij}=+1$, full lines) and the horizontal couplings
are random ($J_{ij}=\pm 1$, dashed lines).
\medskip
\item{Fig. 6}
$\pm J$ spin glass in  $2d$:
quenched free energy of the system obtained by a decomposition of the lattice
into independent strips,
   as function of the temperature $T$ at different
 widths $L=1,2,3,8$ of the strips.
\bye